\documentstyle[12pt]{article}

\def\beq{\begin{equation}}
\def\eeq{\end{equation}}
\def\barr{\begin{eqnarray}}
\def\earr{\end{eqnarray}}
\def\dag{\dagger}

\def\tr{{\rm tr}}
\def\l{\left(}
\def\r{\right)}

\begin{document}

\title{Vacuum Mass Spectra for SU(N) Self-Dual Chern-Simons-Higgs
Systems\footnote{UCONN-94-4;hep-th/9408061}}

\author{\normalsize{Gerald Dunne} \\
\normalsize{Department of Physics}\\
\normalsize{University of Connecticut}\\
\normalsize{Storrs, CT 06269 USA}\\
   \\
\normalsize{dunne@hep.phys.uconn.edu} \\}

\date{}

\maketitle

\begin{abstract}
We study the $SU(N)$ self-dual Chern-Simons-Higgs systems with adjoint matter
coupling, and show that the sixth order self-dual potential has $p(N)$ gauge
inequivalent degenerate minima, where $p(N)$ is the number of partitions of
$N$. We compute the masses of the gauge and scalar excitations in these
different vacua, revealing an intricate mass structure which reflects the
self-dual nature of the model.
\end{abstract}

\section{Introduction}

Relativistic self-dual Chern-Simons-Higgs systems in $2+1$ dimensions have been
shown to possess many remarkable properties. In the abelian theories
\cite{hong}, with the scalar potential of a particular sixth order form,
the energy functional is bounded below by a Bogomol'nyi style bound \cite{bog}.
This lower bound is saturated by topological solitons and nontopological
vortices \cite{roman}. Furthermore, the self-dual structure of the
Chern-Simons-Higgs system is related at a fundamental level to an $N=2$
supersymmetry in $2+1$ dimensions \cite{clee,hlousek,kaolee}. The self-dual
structure of these abelian Chern-Simons-Higgs systems has been shown to extend
to nonabelian Chern-Simons-Higgs systems with a global $U(1)$ symmetry
\cite{klee1}, once again with a special sixth order scalar potential. However,
while the self-dual structure generalizes in a relatively straightforward
manner, the analysis of the nonabelian self-duality equations themselves is
significantly more complicated. This complication is further compounded by the
many different choices: of gauge group, of representation, of matter coupling,
etc... . Matter fields in the defining representation were studied in
\cite{klee2}, while the adjoint matter coupling, in which one can treat the
gauge and matter fields in the same representation, has been studied in
\cite{klee1,dunne1}. Recently, the $SU(3)$ self-dual Chern-Simons-Higgs system
with adjoint coupling has been investigated in detail, with a systematic
analysis of the three distinct degenerate vacua \cite{kao}. In this paper, I
present the mass spectra of the various gauge inequivalent degenerate vacua of
the $SU(N)$ self-dual Chern-Simons-Higgs system with adjoint coupling. While I
concentrate on $SU(N)$, the approach is readily generalizable to other compact
gauge groups. In the $SU(N)$ case, the number of gauge inequivalent minima is
equal to the number, $p(N)$, of partitions of $N$. The mass spectra in these
vacua reveal a remarkably intricate structure, reflecting the self-duality
symmetry in conjunction with the Chern-Simons Higgs mechanism \cite{deser}.

In Section 2 I introduce the model and briefly review the derivation of the
relativistic self-dual Chern-Simons equations. The potential minima are found
by solving an algebraic embedding problem, and this fact is exploited in
Section 3 to provide a complete and constructive classification of the gauge
inequivalent degenerate vacua. In Section 4 I analyze the masses of the gauge
and scalar excitations in these various vacua, and discuss some of the
interesting features which arise. The most involved nontrivial vacuum is the
maximal symmetry breaking case, for which the complete $SU(N)$ mass spectrum
is presented. The mass matrices of the real fields are discussed in Section 5.
Section 6 is devoted to some conclusions and suggestions for further
investigation.

\section{Relativistic Self-Duality Equations}

Consider the following Lagrange density in $2+1$ dimsensional spacetime
\beq
{\cal L}=-\tr\left(\left(D_\mu \phi\right)^\dag D^\mu \phi\right) -\kappa
\epsilon^{\mu \nu \rho} \tr\left(\partial_\mu A_\nu A_\rho +{2\over 3}
A_\mu A_\nu A_\rho \right) - V\left(\phi, \phi^\dag\right)
\label{lag}
\eeq
where the gauge invariant scalar field potential $V(\phi, \phi^\dag)$ is
\beq
V\l\phi, \phi^\dag\r = {1\over 4\kappa^2}\tr\l \l[\;[\;\phi,
\phi^\dag\;],\phi\;]-v^2\phi\r^\dag\;\l[\;[\;\phi, \phi^\dag\;],\phi\;]-
v^2\phi\r\r.
\label{pot}
\eeq
The covariant derivative is $D_\mu \equiv\partial_\mu+[A_\mu,~\;]$, the
space-time metric is taken to be $g_{\mu\nu}={\rm diag}\l -1,1,1 \r$, and
$\tr$ refers to the trace in a finite dimensional representation of the
compact simple Lie algebra ${\cal G}$ to which the gauge fields $A_\mu$
and the charged matter fields $\phi$ and $\phi^\dag$ belong. Most of the
discussion will focus on the Lie algebra of $SU(N)$, but the generalization to
an arbitrary compact simple Lie algebra is straightforward and is indicated
at the appropriate points. The $v^2$ parameter appearing in the potential
(\ref{pot}) plays the role of a mass parameter.

The Euler-Lagrange equations of motion obtained from the Lagrange density
(\ref{lag}) are
\barr
D_\mu D^\mu \phi&=&{\partial V \over \partial \phi^\dag}
\label{phieq}
\earr
\barr
- \kappa \epsilon^{\mu \nu \rho}F_{\nu \rho}& = &i J^{\mu}
\label{aeq}
\earr
where $F_{\nu \rho}\equiv\partial_\nu A_\rho -\partial_\rho A_\nu
+[\;A_\nu, A_\rho\;]$ is the gauge curvature, and the  nonabelian current
$J^\mu$ is given by
\beq
J^\mu \equiv -i\l[\;\phi^\dag, D^\mu \phi\;] - [\;(D^\mu\phi)^\dag,
\phi\;]\r
\label{current}
\eeq
Note that the currrent $J_\mu$ is covariantly conserved, $D_\mu J^\mu=0$,
while the gauge invariant current
\beq
V^\mu= -i \tr\l\phi^\dag D^\mu \phi - \l D^\mu \phi\r^\dag \phi\r
\label{abeliancurrent}
\eeq
is ordinarily conserved: $\partial_\mu V^\mu =0$.

The energy density for this system can be expressed as
\cite{klee1,dunne1,kao}
\barr
{\cal E} &=& \tr\l\l D_0\phi-{i\over 2\kappa}\l[\;[\;\phi,\phi^\dag\;],\phi\;]-
v^2 \phi\r\r^\dag\l D_0\phi-{i\over
2\kappa}\l[\;[\;\phi,\phi^\dag\;],\phi\;]-
 v^2 \phi\r\r\r\cr\cr
& & + \tr\l\l D_-\phi\r^\dag D_-\phi\r +{iv^2 \over 2\kappa}\tr \l \phi^\dag
\l D_0\phi\r -\l D_0 \phi\r^\dag \phi \r
\label{energy}
\earr
where $D_-=D_1-iD_2$, and $\kappa$ has been chosen positive. The first two
terms in (\ref{energy}) are manifestly positive and the third gives a lower
bound for the energy density, which may be written in terms of the time
component, $V^0$, of the gauge invariant current defined in
(\ref{abeliancurrent}):
\beq
{\cal E} \geq {v^2 \over 2\kappa} V^0
\label{bound}
\eeq
This lower bound is saturated when the following two conditions (each first
order in spacetime derivatives) hold:
\barr
D_- \phi&=&0\\
\label{condition1}
D_0 \phi&=&{i\over 2\kappa} \l [\;[\;\phi,\phi^\dag\;],\phi ]-v^2\phi\r
\label{condition2}
\earr
The consistency condition of these two equations states that
\barr
\l D_0 D_- -D_- D_0 \r \phi &\equiv& [F_{0-},\phi ]\cr
&=&-{i\over 2\kappa} [[ \phi , (D_+ \phi )^\dag ], \phi ]\cr
&=&{1\over 2 \kappa} [ J_- , \phi ]
\earr
which expresses the gauge field Euler-Lagrange equation of motion,
$F_{0-}={1\over 2\kappa}J_-$, for the spatial component of the current. The
other gauge field equation, $F_{+-}={1\over \kappa} J_0$, may be re-expressed,
using equation (\ref{condition2}), in a form not involving explicit time
derivatives. We thus arrive at the ``{\it relativistic self-dual Chern-Simons
equations}'':
\barr
D_- \phi &=&0
\label{sd1}
\earr
\barr
F_{+-} &=& {1\over \kappa^2} [ v^2 \phi -[  [ \phi, \phi^\dag ], \phi]
,\phi^\dag  ]
\label{sd2}
\earr
Note that if we ignore the quartic $\phi$ term in (\ref{sd2}), which
corresponds to taking the nonrelativistic limit \cite{dunne1}, then these
relativistic self-duality equations (\ref{sd1}, \ref{sd2}) reduce to the
nonrelativistic self-dual Chern-Simons equations studied in
\cite{grossman,djpt,dunne2}.

At the self-dual point, we can use equation (\ref{condition2}) to express the
energy density as
\beq
{\cal E}_{\rm SD} = {v^2\over 2\kappa^2} \tr\l\phi^\dag \l v^2 \phi-[  [ \phi,
\phi^\dag ], \phi] \r\r
\label{newbound}
\eeq
Recall that all solutions to the {\bf nonrelativistic} self-duality equations
correspond to the static zero-energy solutions to the Euler-Lagrange
equations of motion \cite{djpt}. Here, in the relativistic theory, the
situation is rather different. First, the lower bound (\ref{bound}) on the
energy density is not necessarily zero, and the solutions of (\ref{condition2})
are time dependent. Furthermore, unlike in the nonrelativistic case, it is
possible to have nontrivial solutions for $\phi$ while having $F_{+-}=0$.
These solutions do have zero energy, and are gauge equivalent to solutions of
the {\it algebraic} equation
\beq
[ [ \phi, \phi^\dag ], \phi ] =v^2 \phi .
\label{min}
\eeq
Solutions of this equation also correspond to the minima of the potential
(\ref{pot}), and these potential minima are clearly degenerate.

A class of solutions to the self-duality equations (\ref{sd2}) is given by the
following zero energy solutions of the Euler-Lagrange equations:
\barr
\phi&=&g^{-1} \phi_{(0)} g  \cr
A_\pm &=& g^{-1} \partial_\pm g \cr
A_0&=&g^{-1} \partial_0 g
\label{zeroenergy}
\earr
where $\phi_{(0)}$ is any solution of (\ref{min}), and $g=g(\vec{x},t)$ takes
values in the gauge group. It is clear that these solutions satisfy $D_0
\phi=0$, $D_-\phi=0$, $F_{+-}=0$, as well as the algebraic equation
(\ref{min}), which implies that they are self-dual, and that they have zero
magnetic field and zero charge density.

While this class of solutions looks somewhat trivial, it is still important
because the solutions, $\phi_{(0)}$, of the algebraic equation (\ref{min})
classify the minima of the potential $V$, and the finite {\it nonzero} energy
solutions of the self-duality equations must be gauge equivalent to such a
solution at infinity:
\beq
\phi \to g^{-1} \phi_{(0)} g \quad\quad {\rm as} \quad r\to\infty
\eeq

\section{Classification of Minima}

As has been pointed out in the context of the $SU(2)$ and $SU(3)$ models
\cite{klee1,kao}, equation (\ref{min}) is just the $SU(2)$ commutation
relation, once a factor of $v$ has been absorbed into the field $\phi$. For a
general gauge algebra, finding the solutions to (\ref{min}) is the classic
Dynkin problem \cite{dynkin} of embedding $SU(2)$ into a general Lie
algebra.\footnote{It is interesing to note that this type of embedding problem
also plays a significant role in the theory of spherically symmetric magnetic
monopoles and the Toda molecule equations \cite{leznov}.}

It is clear that in order to satisfy (\ref{min}) for a general gauge algebra,
$\phi=\phi_{(0)}$ must be a linear combination of the step operators for the
{\it positive} roots of the algebra. Further, since we have the freedom of
global gauge invariance, we can choose representative gauge inequivalent
solutions $\phi_{(0)}$ to be linear combinations of the step operators of the
positive {\it simple} roots. It is therefore convenient to work in the
Chevalley basis \cite{humphreys} for the gauge algebra (for ease of
presentation we shall present formulas for the simply-laced algebras). In the
Chevalley basis, the Cartan subalgebra elements, $H_a$, and the simple root
step operators, $E_a$, have the following simple commutation relations
\barr
[H_a , H_b ]&=&0\cr
[E_a, E_{-b} ] &=&\delta_{a\, b} H_a\cr
[H_a , E_{\pm b}]&=&\pm K_{b\, a}E_{\pm b}
\label{commutators}
\earr
where $a$ and $b$ take values $1\dots r$ ($r$ is the rank of the
algebra), and $K_{a\, b}$ is the Cartan matrix which encapsulates the inner
products of the simple roots $\vec{\alpha}^{(a)}$:
\barr
K_{a\, b} \equiv
2\; {\vec{\alpha}^{(a)}\cdot\vec{\alpha}^{(b)}\over\vec{\alpha}^{(b)}\cdot
\vec{\alpha}^{(b)}}
\label{cartan}
\earr
The step operators satisfy $E_{-a}=E_a^\dag$, and the generators are
normalized in the Chevalley basis as:
\barr
\tr \l H_a H_b \r &=& K_{a\, b}\cr
\tr \l E_a E_{-b} \r &=& \delta_{a\, b}\cr
\tr \l H_a E_{\pm b} \r &=&0
\label{traces}
\earr
In this paper we concentrate on the gauge algebra $SU(N)$, for which the
Cartan
matrix $K$ is the $(N-1)\times (N-1)$ symmetric tridiagonal matrix:
\barr
K=\left(\matrix{2&-1&0&\dots&&0\cr -1&2&-1&0&\cr
0&-1&2&-1&0\cr \vdots&&&&&\vdots\cr
0&\dots&&0&-1&2}\right)
\earr
Expand
$\phi_{(0)}$ in terms of the positive simple root step operators as:
\beq
\phi_{(0)} = \sum_{a=1}^{N-1} \phi_{(0)}^a E_a
\label{expansion}
\eeq
Then $[\phi_{(0)}, \phi_{(0)}^\dag ]$ is diagonal,
\beq
[\phi_{(0)}, \phi_{(0)}^\dag ] = \sum_{a=1}^{N-1} |\phi_{(0)}^a |^2 H_a .
\eeq
The commutation relations in (\ref{commutators}) then imply that
\barr
[[\phi_{(0)}, \phi_{(0)}^\dag ], \phi_{(0)}] = \sum_{a=1}^{N-1}
\sum_{b=1}^{N-1} |\phi_{(0)}^a |^2 \phi_{(0)}^b K_{b\, a}E_b
\earr
which, like $\phi_{(0)}$, is once again a linear combination of just the simple
root step operators. Thus, for suitable choices of the coefficients
$\phi_{(0)}^a$ it is possible for the $SU(N)$ algebra element $\phi_{(0)}$ to
satisfy the $SU(2)$ commutation relation $[ [ \phi, \phi^\dag ], \phi ]
=\phi$.

For example, one can always choose $\phi_{(0)}$ proportional to a {\it
single} step operator, which by global gauge invariance can always be taken to
be $E_1$ :
\barr
\phi_{(0)}={1\over \sqrt{2}} E_1
\label{first}
\earr
In the other extreme, the $SU(N)$ ``maximal embedding'' case, with {\it all}
$N-1$ step operators involved in the expansion (\ref{expansion}), the solution
for $\phi_{(0)}$ is :\footnote{In general, the squares of the coefficients are
the coefficients, in the simple root basis, of (one half times) the sum of {\it
all} positive roots of the algebra.}
\beq
\phi_{(0)}={1\over \sqrt{2}}\sum_{a=1}^{N-1} \sqrt{a(N-a)}\; E_a
\label{max}
\eeq
All other solutions for $\phi_{(0)}$, intermediate between the two extremes
(\ref{first}) and (\ref{max}), can be generated by the following systematic
procedure. If one of the simple root step operators, say $E_b$, is omitted from
the summation in (\ref{expansion}) then this effectively decouples the $E_{\pm
a}$'s with $a<b$ from those with $a>b$. Then the coefficients for the $(b-1)$
step operators $E_a$ with $a<b$ are just those for the maximal embedding
(see equation (\ref{max})) in $SU(b)$, and the coefficients for the $(N-b-1)$
$E_a$'s with $a>b$ are those for the maximal embedding in $SU(N-b)$:
\barr
\phi_{(0)}={1\over \sqrt{2}}\sum_{a=1}^{b-1} \sqrt{a(b-a)} E_a +{1\over
\sqrt{2}}\sum_{a=b+1}^{N-1} \sqrt{a(N-b-a)} E_a
\label{oneomission}
\earr
Diagrammatically, we can represent the maximal embedding case
(\ref{max}) with the Dynkin diagram of $SU(N)$ :
\barr
\underbrace{o-o-o- \dots -o-o}_{N-1}
\earr
which shows the $N-1$ simple roots of the algebra, each connected to its
nearest neighbours by a single line. Omitting the $b^{th}$ simple root step
operator from the sum in (\ref{expansion}) can be conveniently represented
as breaking the Dynkin diagram in two by deleting the $b^{th}$ dot:
\barr
\underbrace{o-o- \dots -o}_{b-1} \;\; \times \;\; \underbrace{o- \dots
-o}_{N-b-1}
\earr
With this deletion of the $b^{th}$ dot, the $SU(N)$ Dynkin diagram breaks
into the Dynkin diagram for $SU(b)$ and that for $SU(N-b)$. Since the
remaining simple root step operators decouple into a Chevalley basis for
$SU(b)$ and another for $SU(N- b)$, the coefficients required for the
summation over the first $b-1$ step  operators are just those given in
(\ref{max}) for the maximal embedding in $SU(b)$, while the coefficients for
the summation over the last $N-b-1$ step operators are given by the
maximal embedding for $SU(N-b)$, as indicated in (\ref{oneomission}).

It is clear that this process may be repeated with further roots being
deleted from the Dynkin diagram, thereby subdividing the original $SU(N)$
Dynkin diagram, with  its $N-1$ consecutively linked dots, into subdiagrams of
$\leq N-1$ consecutively linked dots. The final diagram, with $M$ deletions
made, can be characterized, up to gauge equivalence, by the $M+1$ lengths of
the remaining consecutive strings of dots. A simple counting argument shows
that the total number of ways of doing this (including the case where {\it all}
dots are deleted, which corresponds to the trivial solution $\phi_{(0)}=0$) is
given by the number, $p(N)$, of (unrestricted) partitions of $N$.

The $SU(4)$ case is sufficient to illustrate this procedure. There are 5
partitions of 4, and they correspond to the following solutions for
$\phi_{(0)}$:
\barr
o-o-o {\hskip 1in}&& \phi_{(0)}={1\over \sqrt{2}}\l \sqrt{3}E_1
+2E_2+\sqrt{3}E_3
\r\cr
o-o\;\;\times {\hskip 1in}&& \phi_{(0)}=E_1+E_2\cr
o\;\;\;\times\;\;\; o {\hskip 1in}&& \phi_{(0)}={1\over \sqrt{2}} E_1
+{1\over
\sqrt{2}} E_3 \cr
o\;\;\;\times\;\;\;\times {\hskip 1in}&&\phi_{(0)}={1\over \sqrt{2}}E_1\cr
\times\;\;\;\times\;\;\;\times {\hskip 1in}&& \phi_{(0)}=0
\label{su4example}
\earr

Thus we have a simple constructive procedure, and a correspondingly simple
labelling notation, for finding all $p(N)$ gauge inequivalent solutions
$\phi_{(0)}$ to the algebraic embedding condition (\ref{min}). Recall that each
such $\phi_{(0)}$ characterizes a distinct minimum of the potential $V$, as
well as a class of zero energy solutions to the selfduality equations
(\ref{sd2}).

Since each vacuum solution $\phi_{(0)}$ corresponds to an embedding of
$SU(2)$ into $SU(N)$, an alternative shorthand for labelling the different
vacua consists of listing the block diagonal spin content of the $SU(2)$ Cartan
subagebra element $[\phi_{(0)}, \phi_{(0)}^\dag ]\sim J_3$. For example,
consider the matter fields $\phi$ taking values in the $N\times N$ defining
representation. Then, for each vacuum solution, $[\phi_{(0)},\phi_{(0)}^\dag ]$
takes the $N\times N$ diagonal sub-blocked form:
\barr
[\phi_{(0)}, \phi_{(0)}^\dag ]=\left( \matrix{j_1\cr &\ddots \cr &&-j_1\cr
&&&j_2\cr &&&&\ddots\cr &&&&&-j_2\cr &&&&&&\ddots\cr
&&&&&&&j_M\cr
&&&&&&&&\ddots\cr &&&&&&&&&-j_M}\right)
\earr
Each spin $j$ sub-block has dimension $2j+1$, and so it is therefore natural
to associate this particular $\phi_{(0)}$ with the following partition of $N$ :
\barr
N=(2j_1+1)+(2j_2+1)+\dots +(2j_M+1)
\earr
For example, the $SU(4)$ solutions listed in (\ref{su4example}) may be
labelled by the partitions $4$, $3+1$, $2+2$, $2+1+1$, and $1+1+1+1$,
respectively.

\section{Vacuum Mass Spectra}

Having classified all possible gauge inequivalent vacua of the potential
$V$, we now determine the spectrum of massive excitations in each vacuum. In
the abelian model \cite{hong,roman} there is only one nontrivial vacuum, and
a consequence of the particular $6^{th}$ order self-dual form of the potential
was that in this broken vacuum the massive gauge excitation and the
remaining real massive scalar field were degenerate in mass.\footnote{This
degeneracy of the gauge and scalar masses in the broken vacuum is also true of
the $2+1$ dimensional Abelian Higgs model \cite{bog}.} In the nonabelian models
considered here the situation is considerably more complicated, due to the
presence of many fields and also due to the many different gauge inequivalent
vacua. Nevertheless, we shall see that an analogous mass degeneracy pattern
exists.

The scalar masses are determined by expanding the shifted potential
$V(\phi+\phi_{(0)})$ to quadratic order in the field $\phi$:
\barr
V(\phi +\phi_{(0)})&=&{v^4\over 4\kappa^2}\tr\l | [[\phi_{(0)},\phi^\dag
],\phi_{(0)}] + [[\phi, \phi_{(0)}^\dag ], \phi_{(0)} ]+[[\phi_{(0)},
\phi_{(0)}^\dag ], \phi ] - \phi |^2  \r
\label{scalarmasses}
\earr
With the fields normalized appropriately, the masses are then given by the
square roots of the eigenvalues of the $2(N^2-1)\times 2(N^2-1)$ mass matrix in
(\ref{scalarmasses}). Since $V$ is a $6^{th}$ order potential\footnote{Note
that a $6^{th}$ order potential is renormalizable in three dimensional
spacetime.}, diagonalizing this scalar field mass matrix is considerably more
complicated than for the conventional $\phi^4$ Higgs model.

In the unbroken vacuum, with $\phi_{(0)}=0$, there are $N^2-1$ complex
scalar fields, each with mass
\barr
m={v^2\over 2\kappa}
\label{mass}
\earr
In one of the broken vacua, where $\phi_{(0)}\neq 0$, some of these $2(N^2-
1)$ massive scalar degrees of freedom are converted to massive gauge
degrees of freedom. The gauge masses are determined by expanding $\tr\l
\l D_\mu\l\phi+\phi_{(0)}\r\r ^\dag \l D_\mu \l\phi+\phi_{(0)}\r\r\r$ and
extracting the piece quadratic in the gauge field $A$:
\barr
v^2 \; \tr\l [A_\mu ,\phi_{(0)}]^\dag [A^\mu , \phi_{(0)} ]\r
\label{gaugemasses}
\earr
Since the Lagrange density (\ref{lag}) for this model only contains a
Chern-Simons term for the gauge fields, and no Yang-Mills term, the gauge
field masses are generated by the Chern-Simons-Higgs mechanism
\cite{deser,dunne3}, which is different from the conventional Higgs
mechanism. Because the Chern-Simons term is {\it first order} in spacetime
derivatives, a quadratic term $v^2 A_\mu A^\mu$ coming from one algebraic
component of (\ref{gaugemasses}) leads to a gauge mode of mass $\sim v^2$ (and
not $\sim v$ as would be the case in the conventional Higgs mechanism). Thus,
the gauge masses are determined by finding the eigenvalues ({\it not} the
square roots of the eigenvalues) of the $(N^2-1)\times (N^2-1)$ mass matrix in
(\ref{gaugemasses}).

This procedure of finding the eigenvalues of the scalar and gauge mass
matrices, must be performed for each of the $p(N)$ gauge inequivalent
minima $\phi_{(0)}$ of $V$. The results for $SU(3)$, $SU(4)$ and $SU(5)$ are
presented in Tables \ref{su3}, \ref{su4} and \ref{su5}. The masses for the two
nontrivial vacua in $SU(3)$ are in agreement with the results of \cite{kao}.

\begin{table}[b]
\center
\begin{tabular}{|c|cc|cccc|} \hline
\multicolumn{1}{|c|}{\rm vacuum }&\multicolumn{6}{c|}{\rm gauge
masses}\\ \cline{2-7}
\multicolumn{1}{|c|}{\rm $\phi_{(0)}$}&\multicolumn{2}{c|} {\rm real}&
\multicolumn{4}{c|}{\rm complex}\\
\multicolumn{1}{|c|}{\rm }&\multicolumn{2}{c|}{\rm fields}&
\multicolumn{4}{c|}{\rm fields}\\ \hline\hline
${1\over \sqrt{2}}E_1$&2&&1/2&1/2&1& \\ \hline
$E_1+E_2$&2&6&1&2&5& \\ \hline
\multicolumn{7}{c}{\rm } \\ \hline
\multicolumn{1}{|c|}{\rm vacuum }&\multicolumn{6}{c|}{\rm scalar
masses}\\ \cline{2-7}
\multicolumn{1}{|c|}{\rm $\phi_{(0)}$}&\multicolumn{2}{|c|} {\rm real}&
\multicolumn{4}{c|}{\rm complex}\\
\multicolumn{1}{|c|}{\rm }&\multicolumn{2}{c|}{\rm fields}&
\multicolumn{4}{c|}{\rm fields}\\ \hline\hline
${1\over \sqrt{2}}E_1$&2&&1&3/2&3/2&2 \\ \hline
$E_1+E_2$&2&6&2&3&5& \\ \hline
\end{tabular} \\
\caption{$SU(3)$ vacuum mass spectra, in units of the fundamental mass scale
${v^2\over 2\kappa}$, for the inequivalent nontrivial minima $\phi_{(0)}$ of
the potential $V$. Notice that for each vacuum the {\it total} number of
massive degrees of freedom is equal to $2(N^2-1)=16$, although the distribution
between gauge and scalar fields is vacuum dependent.}
\label{su3}
\end{table}

\begin{table}[b]
\center
\begin{tabular}{|c|ccc|ccccccccc|} \hline
\multicolumn{1}{|c|}{\rm vacuum }&\multicolumn{12}{c|}{\rm gauge
masses}\\ \cline{2-13}
\multicolumn{1}{|c|}{\rm $\phi_{(0)}$}&\multicolumn{3}{|c|} {\rm real}&
\multicolumn{9}{c|}{\rm complex}\\
\multicolumn{1}{|c|}{\rm }&\multicolumn{3}{c|}{\rm fields}&
\multicolumn{9}{c|}{\rm fields}\\ \hline\hline
${1\over \sqrt{2}}E_1$&2&&&1/2&1/2&1/2&1/2&1&&&& \\ \hline
${1\over \sqrt{2}}E_1+{1\over \sqrt{2}}E_3$&2&2&&1&1&1&1&2&&&& \\ \hline
$E_1+E_2$&2&6&&1&1&1&2&2&5&&& \\ \hline
${1\over \sqrt{2}}\l
\sqrt{3}E_1+2E_2+\sqrt{3}E_3\r$&2&6&12&1&2&3&5&8&11&&& \\ \hline
\multicolumn{7}{c}{\rm } \\ \hline
\multicolumn{1}{|c|}{\rm vacuum }&\multicolumn{12}{c|}{\rm scalar
masses}\\ \cline{2-13}
\multicolumn{1}{|c|}{\rm $\phi_{(0)}$}&\multicolumn{3}{|c|} {\rm real}&
\multicolumn{9}{c|}{\rm complex}\\
\multicolumn{1}{|c|}{\rm }&\multicolumn{3}{c|}{\rm fields}&
\multicolumn{9}{c|}{\rm fields}\\ \hline\hline
${1\over \sqrt{2}}E_1$&2&&&1&1&1&1&3/2&3/2&3/2&3/2&2 \\ \hline
${1\over \sqrt{2}}E_1+{1\over \sqrt{2}}E_3$&2&2&&1&1&1&2&2&2&2&2& \\ \hline
$E_1+E_2$&2&6&&1&2&2&2&2&3&5&& \\ \hline
${1\over \sqrt{2}}\l\sqrt{3}E_1+2E_2+\sqrt{3}E_3\r$&2&6&12&2&3&4&5&8&11&&& \\
\hline
\end{tabular} \\
\caption{$SU(4)$ vacuum mass spectra, in units of the fundamental mass scale
${v^2\over 2\kappa}$, for the inequivalent nontrivial minima $\phi_{(0)}$ of
the potential $V$. Notice that for each vacuum the {\it total} number of
massive degrees of freedom is equal to $2(N^2-1)=30$, although the distribution
between gauge and scalar fields is vacuum dependent.}
\label{su4}
\end{table}

\begin{table}[b]
\center
\begin{tabular}{|c|cccc|cccccccc|} \hline
\multicolumn{1}{|c|}{\rm vacuum }&\multicolumn{12}{c|}{\rm gauge
masses}\\ \cline{2-13}
\multicolumn{1}{|c|}{\rm $\phi_{(0)}$}&\multicolumn{4}{|c|} {\rm real}&
\multicolumn{8}{c|}{\rm complex}\\
\multicolumn{1}{|c|}{\rm }&\multicolumn{4}{c|}{\rm fields}&
\multicolumn{8}{c|}{\rm fields}\\ \hline\hline
${1\over \sqrt{2}}E_1$&2&&&&1/2&1/2&1/2&1/2&1/2&1/2&1& \\ \hline
${1\over \sqrt{2}}E_1+{1\over \sqrt{2}}E_3$&2&2&&&1/2&1/2&1/2&1/2&1&1&1&1\\
&&&&&2&&&&&&&\\ \hline
$E_1+E_2$&2&6&&&1&1&1&1&1&2&2&2 \\
&&&&&5&&&&&&&\\ \hline
${1\over \sqrt{2}}E_1+E_3+E_4$&2&2&6&&1/2&1/2&1&1&3/2&3/2&2&5 \\
&&&&&7/2&7/2&&&&&&\\ \hline
${1\over
\sqrt{2}}\l\sqrt{3}E_1+2E_2+\sqrt{3}E_3\r$&2&6&12&&1&3/2&3/2&2&3&7/2&7/2&5\\
&&&&&8&11&&&&&&\\ \hline
$\sqrt{2}E_1+\sqrt{3}(E_2+E_3)+\sqrt{2}E_4$&2&6&12&20&1&2&3&4&5&8&11&11\\
&&&&&16&19&&&&&&\\ \hline
\multicolumn{7}{c}{\rm } \\ \hline
\multicolumn{1}{|c|}{\rm vacuum }&\multicolumn{12}{c|}{\rm scalar
masses}\\ \cline{2-13}
\multicolumn{1}{|c|}{\rm $\phi_{(0)}$}&\multicolumn{4}{|c|} {\rm real}&
\multicolumn{8}{c|}{\rm complex}\\
\multicolumn{1}{|c|}{\rm }&\multicolumn{4}{c|}{\rm fields}&
\multicolumn{8}{c|}{\rm fields}\\ \hline\hline
${1\over \sqrt{2}}E_1$&2&&&&1&1&1&1&1&1&1&1 \\
&&&&&1&3/2&3/2&3/2&3/2&3/2&3/2&2 \\ \hline
${1\over \sqrt{2}}E_1+{1\over \sqrt{2}}E_3$&2&2&&&1&1&1&1&3/2&3/2&3/2&3/2\\
&&&&&2&2&2&2&2&&&\\ \hline
$E_1+E_2$&2&6&&&1&1&1&1&2&2&2&2 \\
&&&&&2&2&2&3&5&&&\\ \hline
${1\over \sqrt{2}}E_1+E_3+E_4$&2&2&6&&1&3/2&3/2&2&2&5/2&5/2&7/2 \\
&&&&&7/2&3&5&&&&&\\ \hline
${1\over
\sqrt{2}}\l\sqrt{3}E_1+2E_2+\sqrt{3}E_3\r$&2&6&12&&1&2&5/2&5/2&3&7/2&7/2&4\\
&&&&&5&8&11&&&&&\\ \hline
$\sqrt{2}E_1+\sqrt{3}(E_2+E_3)+\sqrt{2}E_4$&2&6&12&20&2&3&4&5&5&8&11&11\\
&&&&&16&19&&&&&&\\ \hline
\end{tabular} \\
\caption{$SU(5)$ vacuum mass spectra, in units of the fundamental mass scale
${v^2\over 2\kappa}$, for the inequivalent nontrivial minima $\phi_{(0)}$ of
the potential $V$. Notice that for each vacuum the {\it total} number of
massive degrees of freedom is equal to $2(N^2-1)=48$, although the distribution
between gauge and scalar fields is vacuum dependent.}
\label{su5}
\end{table}

A number of interesting observations can be made at this point, based on the
evaluation of these mass spectra for the various vacua in $SU(N)$ for $N$ up to
$10$.

(i) All masses, both gauge and scalar, are integer or half-odd-integer
multiples of the fundamental mass scale $m=v^2/2\kappa$. The fact that all
the scalar masses are proportional to $m$ is clear from the form of the
potential $V$ in (\ref{pot}). The fact that the gauge masses are multiples of
the {\it same} mass scale depends on the fact that the Chern-Simons coupling
parameter $\kappa$ has been included in the overall normalization of the
potential in (\ref{pot}). This is a direct consequence of the self-duality of
the model.

(ii) In each vacuum, the masses of the real scalar excitations are equal
to the masses of the real gauge excitations, whereas this is not true of the
complex scalar and gauge fields\footnote{By `complex' gauge fields we simply
mean those fields which naturally appear as complex combinations of the
(nonhermitean) step operator generators.}. Indeed, in some vacua the {\it
number} of complex scalar degrees of freedom and complex gauge degrees of
freedom is not even the same. This will be discussed further in the next
section.

(iii) In each vacuum, each mass appears at least twice, and always an
even number of times. For the complex fields this is a triviality, but for the
real fields this is only true as a consequence of the feature mentioned in
(ii). This pairing of the masses is a reflection of the $N=2$ supersymmetry
of the relativistic self-dual Chern-Simons systems \cite{clee}.

(iv) While the distribution of masses between gauge and scalar modes is
different in the different vacua, the total number of degrees of freedom is, in
each case, equal to $2(N^2-1)$, as in the unbroken phase.

The most complicated, and most interesting, of the nontrivial vacua is the
``maximal embedding'' case, with $\phi_{(0)}$ given by (\ref{max}). For this
vacuum, the gauge and scalar mass spectra have additional features of note.
First, this ``maximal embedding'' also corresponds to ``maximal symmetry
breaking'', in the sense that in this vacuum all $N^2-1$ gauge degrees of
freedom acquire a mass. The original $2(N^2-1)$ massive scalar modes
divide equally between the scalar and gauge fields. Moreover, the mass
spectrum reveals an intriguing and intricate pattern, as shown in Table
\ref{sun}. It is interesting to note that for the $SU(N)$ maximal symmetry
breaking vacuum, the entire scalar mass spectrum is {\it almost} degenerate
with the gauge mass spectrum : there is just {\it one} single complex
component for which the masses differ!

\begin{table}[b]
\center
\begin{tabular}{|c|ccccccc|} \hline
\multicolumn{8}{|c|}{\rm gauge masses}\\ \hline
\multicolumn{1}{|c|}{\rm real}&\multicolumn{7}{c|}{\rm complex} \\
\multicolumn{1}{|c|}{\rm fields}&\multicolumn{7}{c|}{\rm fields}\\
\hline
2&1&2&3&4&5&\dots&N-1  \\
6&5&8&11&14&\dots&3N-4& \\
12&11&16&21&\dots&5N-9&&\\
20&19&26&\dots&7N-16&&&\\
30&29&\dots&9N-25&&&& \\
\vdots&\vdots&&&&&&\\
N(N-1)&N(N-1)-1&&&&&&\\ \hline
\multicolumn{8}{c}{\rm } \\ \hline
\multicolumn{8}{|c|}{\rm scalar masses} \\ \hline
\multicolumn{1}{|c|}{\rm real}&\multicolumn{7}{c|}{\rm complex}\\
\multicolumn{1}{|c|}{\rm fields}&\multicolumn{7}{c|}{\rm fields}\\
\hline
2&N&2&3&4&5&\dots&N-1 \\
6&5&8&11&14&\dots&3N-4& \\
12&11&16&21&\dots&5N-9&&   \\
20&19&26&\dots&7N-16&&& \\
30&29&\dots&9N-25&&&& \\
\vdots&\vdots&&&&&& \\
N(N-1)&N(N-1)-1&&&&&&  \\ \hline
\end{tabular} \\
\caption{$SU(N)$ mass spectrum, in units of the fundamental mass scale
${v^2\over 2\kappa}$, for the maximal symmetry breaking vacuum, for which
$\phi_{(0)}$ is given by (\protect{\ref{max}}). Notice that the gauge mass
spectrum and the scalar mass are {\it almost} degenerate - they differ in just
one complex field component.}
\label{sun}
\end{table}

\section{Mass Matrices for Real Fields}

The masses of the real fields exhibit special simple properties, which we
discuss in this section. As mentioned above, in each vacuum $\phi_{(0)}$ the
{\it number} of real scalar modes is equal to the number of real gauge modes.
Furthermore, the two mass spectra coincide exactly, and are all {\it integer}
multiples of the mass scale $m$ in (\ref{mass}). The real gauge fields come
from the diagonal algebraic components $H_a$, while the real scalar fields come
from the simple root step operator components $E_a$. Indeed, the real scalar
fields correspond to those fields shifted by the symmetry breaking minimum
field $\phi_{(0)}$, which is decomposed in terms of the simple root step
operators as in (\ref{expansion}). This means that the {\it number} of real
scalars in a given vacuum $\phi_{(0)}$ is given by the number of nonzero
coefficients $\phi_{(0)}^a$ in the decomposition (\ref{expansion}). This can be
seen explicitly for $SU(3)$, $SU(4)$ and $SU(5)$ in the Tables \ref{su3},
\ref{su4} and \ref{su5}. This also serves as an easy count of the number of
real gauge masses. This also means that to determine the mass matrix for the
real gauge fields we can expand $A_\mu$ in terms of the Cartan subalgebra
elements $H_a$ (the other, off-diagonal, algebraic components do not mix with
these ones at quadratic order). In fact, in order to normalize the gauge fields
correctly, it is more convenient to expand the $A_\mu$ in another Cartan
subalgebra basis, $h_a$, for which the traces are orthonormal (in contrast to
the traces (\ref{traces}) in the Chevalley basis which involve the Cartan
matrix) :
\barr
\tr \l h_a h_b \r = \delta_{a\, b}
\label{newtrace}
\earr
Such basis elements, $h_a$, are related to the Chevalley basis elements, $H_a$,
by
\barr
h_a = \sum_{b=1}^r \omega_a^{(b)}\; H_b
\earr
where $\vec{\omega}^{(b)}$ is the $b^{th}$ fundamental weight of the algebra
\cite{humphreys}, satisfying
\barr
\sum_{b=1}^r \omega_a^{(b)} \alpha_c^{(b)} = \delta_{ac}
\label{orthog}
\earr
where $\vec{\alpha}^{(b)}$ is the $b^{th}$ simple root. For $SU(N)$ we can be
more explicit:
\barr
h_a = {1\over \sqrt{a(a+1)}}\sum_{b=1}^a b\, H_b
\earr
The orthogonality relation (\ref{orthog}) means that the correspondence can be
inverted to give
\barr
H_a = \sum_{b=1}^r \alpha_b^{(a)} \; h_b
\earr
The fundamental weights $\vec{\omega}^{(b)}$ and simple roots
$\vec{\alpha}^{(b)}$ are also related by
\barr
\vec{\alpha}^{(a)} = \sum_{b=1}^r K_{b a} \; \vec{\omega}^{(b)}
\earr
These new basis elements have the following commutation relations with the
simple root step operators:
\barr
[ h_a, E_b ] = \alpha_a^{(b)} E_b
\label{newcomm}
\earr
Given the traces in (\ref{newtrace}) and the commutation relations
(\ref{newcomm}), it is now a simple matter to expand the quadratic gauge field
term (\ref{gaugemasses}) to find the following mass matrix:
\barr
{\cal M}_{a b}^{\rm (gauge)} =  2\;m\;\sum_{c=1}^r |\phi_{(0)}^c |^2\,
\alpha_a^{(c)}\,\alpha_b^{(c)}
\label{gaugemass}
\earr
where $m$ is the fundamental mass scale in (\ref{mass}). For the maximal
embedding vacuum (\ref{max}) in $SU(N)$ this leads to a mass matrix
\barr
{\cal M}_{a b}^{\rm (gauge)} =  m\;\sum_{c=1}^{N-1} c(N-c)\,
\alpha_a^{(c)}\,\alpha_b^{(c)}
\label{sungaugemass}
\earr
This matrix has eigenvalues
\barr
2,\;6,\;12,\;20,\; \dots\;,\:N(N-1)
\label{eigenvalues}
\earr
in multiples of $m$. For any vacuum $\phi_{(0)}$ other than the maximal
symmetry breaking one, the mass matrix for the real gauge fields decomposes
into smaller matrices of the same form, according to the particular partition
of the original $SU(N)$ Dynkin diagram, as described in Section 3.

The real scalar field mass matrix can be computed by expanding the $\phi$ field
appearing in (\ref{scalarmasses}) in terms of the positive root step operators.
With such a decomposition for $\phi$, the quadratic term (\ref{scalarmasses})
simplifies considerably to give a mass (squared) matrix
\barr
{\cal M}_{a b}^{\rm (scalar)}= 4\;m^2\; \phi_{(0)}^a \phi_{(0)}^b \sum_{c=1}^r
|\phi_{(0)}^c |^2 \, K_{a c}\, K_{b c}
\earr
where $K$ is the Cartan matrix (\ref{cartan}). For the $SU(N)$ maximal symmetry
breaking vacuum (\ref{max}) this mass matrix is
\barr
{\cal M}_{a b}^{\rm (scalar)}= m^2\;
\sqrt{a\,b\,(N-a)\,(N-b)}\;\sum_{c=1}^{N-1} c\,(N-c)\, K_{a c}\, K_{b c}
\earr
which has eigenvalues
\barr
(2)^2,\;(6)^2,\;(12)^2,\;(20)^2,\;\dots\;,\;(N(N-1))^2
\label{eigs}
\earr
in units of $m^2$. It is interesting to note that the eigenvalues in
(\ref{eigs}) are the squares of the eigenvalues (\ref{eigenvalues}) of ${\cal
M}^{\rm (gauge)}$, even though ${\cal M}^{\rm (scalar)}$ is {\it not} the
square of the matrix ${\cal M}^{\rm (gauge)}$ in this basis. Nevertheless, as
the real scalar masses are given by the square roots of the eigenvalues in
(\ref{eigs}), we see that the real scalar masses do indeed coincide with the
real gauge masses, a consequence of the $N=2$ supersymmetry of the theory.

\section{Conclusion}

In this paper we have analyzed the vacuum structure of the $SU(N)$ self-dual
Chern-Simons-Higgs systems with adjoint coupling. Finding the locations of the
potential minima (which are degenerate) is equivalent to a classic algebraic
embedding problem. A simple explicit construction is given for enumerating and
evaluating the gauge inequivalent minima for any gauge group. For $SU(N)$, the
number of gauge inequivalent minima is equal to $p(N)$, the number of
partitions of $N$. In the nontrivial vacua, the Chern-Simons-Higgs mechanism
generates masses for some of the algebraic components of the gauge field. Both
the number of and the actual mass values of these gauge excitations depend on
which vacuum is being considered. We have analyzed the resulting mass spectra,
for both the gauge and scalar fields, and identified a number of interesting
symmetry properties of these spectra. There is clearly a very rich structure
present in these spectra, some of which can be understood in terms of the
self-duality, and the associated $N=2$ supersymmetry, of these systems.

The picture is by no means as complete as for the corresponding {\it
nonrelativistic} nonabelian self-dual Chern-Simons-matter systems, where a
classification of all finite charge solutions is known \cite{dunne2}, due to a
deep relationship between the nonrelativistic self-duality equations and
integrable models in two dimensions. Ideally, one would like to discover more
about explicit solutions of the relativistic self-duality equations
(\ref{sd1},\ref{sd2}). Some properties of these equations and their possible
solutions for $SU(2)$ and $SU(3)$ have been discussed in \cite{klee1,kao}. An
analysis of the integrability of the {\it abelian} relativistic models
\cite{schiff} suggests that the relativistic nonabelian self-duality equations
are not completely integrable in general. However, it would be very interesting
to learn if they may be integrable in certain special cases, as was found for
the abelian theories \cite{schiff}. Such information about nonzero energy
solutions to the self-duality equations would shed some light on the
quantization of this model and the quantum role of the intricate vacuum
structure.

\vskip 2in

\noindent{\bf Acknowledgements:} This work has been supported in part by
the D.O.E. through grant number DE-FG02-92ER40716.00, and by the University
of Connecticut Research Foundation. I am grateful to K.~Lee for sending me a
preliminary draft of \cite{kao}.


\end{document}